\documentclass[aps,prd,twocolumn,showpacs,superscriptaddress,amsmath,amssymb,nofootinbib]{revtex4-1}
\usepackage{graphicx}
\usepackage{units}
\usepackage{epsfig}
\usepackage{bigstrut}
\usepackage{overpic}
\usepackage{dcolumn}
\usepackage{bm}
\usepackage{color}
\usepackage{amsmath}
\usepackage{amsfonts}
\usepackage{lineno}
\usepackage{xspace}
\usepackage{multirow}
\usepackage{subfigure}
\usepackage{verbatim}
\usepackage{footmisc}
\usepackage{epstopdf}
\usepackage{xcolor}
\usepackage{soul}
\usepackage{makecell}

\usepackage{dcolumn}

\RequirePackage{lineno}

\usepackage[colorlinks,linkcolor=blue,anchorcolor=blue,citecolor=blue]{hyperref}

\hypersetup{hypertex=true,
	colorlinks=true,
	linkcolor=blue,
	anchorcolor=blue,
	citecolor=blue}

\newcommand{\mev}{\,\unit{MeV}}

\newcommand{\mevcc}{\,\unit{MeV}/c^2}
\newcommand{\gev}{\,\unit{GeV}}

\newcommand{\gevcc}{\,\unit{GeV}/c^2}

\newcommand{\ee}{e^{+}e^{-}}

\newcommand{\pip}{\pi^+}
\newcommand{\pim}{\pi^-}

\newcommand{\lambdacp}{\Lambda_{c}^{+}}
\newcommand{\lcp}{\Lambda_{c}^{+}}
\newcommand{\lambdacm}{\bar{\Lambda}{}_{c}^{-}}
\newcommand{\lcm}{\bar{\Lambda}{}_{c}^{-}}

\newcommand{\Ks}{K_{S}^{0}}
\newcommand{\kp}{K^+}
\newcommand{\km}{K^-}
\newcommand{\kst}{K^{*+}}
\newcommand{\lmdkspi}{\lambdacp\to\Lambda\Ks\pip}

\newcommand{\lmdksk}{\lambdacp\to\Lambda\Ks\kp}
\newcommand{\lmdkst}{\lambdacp\to\Lambda\kst}
\newcommand{\mpipi}{M(\pip\pim)}
\newcommand{\mppi}{M(p\pim)}
\newcommand{\mkspi}{M(\Ks\pip)}
\newcommand{\lcptolx}{\lcp\to\Lambda X}
\newcommand{\lcptoksx}{\lcp\to\Ks X}

\newcolumntype{d}[1]{D{.}{.}{#1}}

\begin{document}

	\title{\boldmath  Measurement of the branching fractions of  the decays $\Lambda_{c}^{+}\rightarrow\Lambda K_{S}^{0}K^{+}$, $\Lambda_{c}^{+}\rightarrow\Lambda K_{S}^{0}\pi^{+}$ and $\Lambda_{c}^{+}\rightarrow\Lambda K^{*+}$}
	
	\author{
	\begin{small}
M.~Ablikim$^{1}$, M.~N.~Achasov$^{4,c}$, P.~Adlarson$^{75}$, O.~Afedulidis$^{3}$, X.~C.~Ai$^{80}$, R.~Aliberti$^{35}$, A.~Amoroso$^{74A,74C}$, Q.~An$^{71,58,a}$, Y.~Bai$^{57}$, O.~Bakina$^{36}$
% ...
\\
\vspace{0.2cm}
(BESIII Collaboration)\\
\vspace{0.2cm} {\it
$^{1}$ Institute of High Energy Physics, Beijing 100049, People's Republic of China\\
% ...
$^{3}$ Bochum  Ruhr-University, D-44780 Bochum, Germany\\
$^{4}$ Budker Institute of Nuclear Physics SB RAS (BINP), Novosibirsk 630090, Russia\\
% ...
$^{35}$ Johannes Gutenberg University of Mainz, Johann-Joachim-Becher-Weg 45, D-55099 Mainz, Germany\\
$^{36}$ Joint Institute for Nuclear Research, 141980 Dubna, Moscow region, Russia\\
% ...
$^{57}$ Southeast University, Nanjing 211100, People's Republic of China\\
$^{58}$ State Key Laboratory of Particle Detection and Electronics, Beijing 100049, Hefei 230026, People's Republic of China\\
% ...
$^{71}$ University of Science and Technology of China, Hefei 230026, People's Republic of China\\
% ...
$^{74}$ University of Turin and INFN, (A)University of Turin, I-10125, Turin, Italy; (B)University of Eastern Piedmont, I-15121, Alessandria, Italy; (C)INFN, I-10125, Turin, Italy\\
$^{75}$ Uppsala University, Box 516, SE-75120 Uppsala, Sweden\\
% ...
$^{80}$ Zhengzhou University, Zhengzhou 450001, People's Republic of China\\
\vspace{0.2cm}
$^{a}$ Deceased\\
% ...
$^{c}$ Also at the Novosibirsk State University, Novosibirsk, 630090, Russia\\
% ...
}
\vspace{0.4cm}
\end{small}
}
	
	\vspace{4cm}
	
	\date{\it \small \bf \today}
	
	\begin{abstract}
Studies are performed of the  Cabibbo-favored decay $\lmdksk$  and the singly Cabibbo-suppressed decay $\lmdkspi$, based on a sample of $\ee$ collision data, corresponding to an integrated luminosity of 4.5 fb$^{-1}$,   accumulated at center-of-mass energies between $4599.53\mev$ and $4698.82\mev$ with the BESIII detector.  The decay $\lmdkspi$ is observed for the first time.
The branching fractions of $\lmdksk$ and  $\lmdkspi$ are measured to be $(3.04\pm0.30\pm0.16)\times 10^{-3}$ and $(1.73\pm0.27\pm0.10)\times 10^{-3}$,  respectively, where the first uncertainties are statistical and the second are systematic.   These results correspond to the most precise measurement of these quantities for both decays.
Evidence of a $\kst$ contribution  in the $\lmdkspi$ decay is found with a statistical significance of $4.7\sigma$. The branching fraction of $\lmdkst$ is calculated under three possible interference scenarios.
	\end{abstract}
	
	\maketitle
	
	
\section{Introduction}
	In contrast to the significant achievements made over the last 20 years in the experimental and theoretical  studies of weak decays of heavy mesons, progress in the area of heavy baryons has been relatively slow~\cite{klein:1990}. The well-known factorization method that has been successfully applied in the study of heavy mesons does not apply to  heavy baryons due to the complexity of the three-quark system~\cite{tseng:1992}. Experimental studies of the decays of charmed baryons provide invaluable information concerning the role of the strong and weak interactions in charm physics. Since its first observation at the Mark~II experiment in 1979~\cite{abrams:1980}, extensive studies have been performed of the $\lcp$, which is the   ground-state charmed baryon. Inclusive measurements yield $\mathcal{B}(\lcptolx)=(38.2^{+2.8}_{-2.2}\pm0.9)\%$~\cite{BESIII:lc2lmdx} and $\mathcal{B}(\lcptoksx)=(9.9\pm0.6\pm0.4)\%$~\cite{BESIII:2020cpu}. However, the summed branching fractions~(BFs) of the known exclusive $\lcp$ decays involving $\Lambda$ and $\Ks$ in the final states are only $(30.4\pm1.3)\%$~\cite{pdg2022} and $(8.1\pm0.4)\%$~\cite{BESIII:2020cpu}, respectively. The difference between the inclusive and summed exclusive results indicate that there is still large room for unknown decays to be discovered.
	
	The decays of the $\lcp$  are dominated by the $c \to s$ transition. Decays that contain one strange hadron have been intensively investigated~\cite{pdg2022}, while $\lcp$ decays into a $\Lambda$ accompanied by at least one strange hadron are theoretically predicted~\cite{Geng:2018upx,Zhao:2018zcb} but have been less studied experimentally~\cite{pdg2022}.  The decays of interest include $\lmdksk$ and $\lmdkspi$.
	
\begin{figure}[!hpbt]
		\centering
		\subfigure[\ Internal $W$-emission]
		{
			\includegraphics[width=0.18\textwidth]{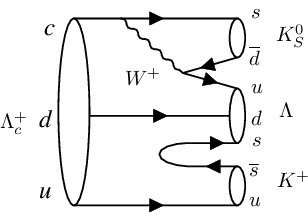}
		}
		\subfigure[\ $W$-exchange]
		{
			\includegraphics[width=0.18\textwidth]{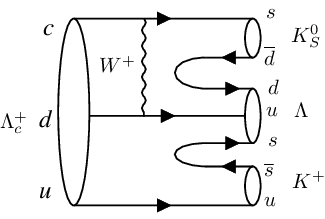}
		}
		\caption{Topology diagrams for $\Lambda_{c}^{+}\to \Lambda K_{S}^{0} K^{+}$.}
		\label{fig:feyaman_lmdksk}
	\end{figure}

\begin{figure}[!hpbt]
		\centering
		\subfigure[\ External $W$-emission]
		{
			\includegraphics[width=0.145\textwidth]{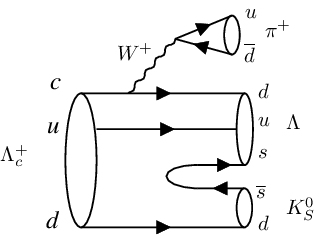}
		}
		\subfigure[\ Internal $W$-emission]
		{
			\includegraphics[width=0.145\textwidth]{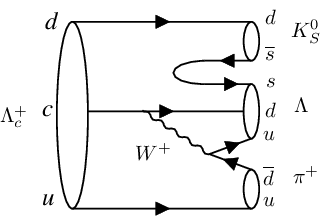}
		}
		\subfigure[\ $W$-exchange]
		{
			\includegraphics[width=0.145\textwidth]{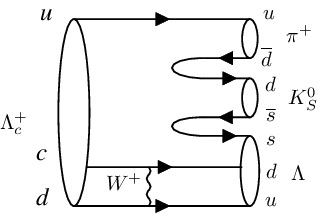}
		}
		\caption{Topology diagrams for $\Lambda_{c}^{+}\to \Lambda K_{S}^{0} \pi^{+}$.}
		\label{fig:feyaman_lmdkspi}
	\end{figure}

\begin{figure}[!hpbt]
		\centering
		\subfigure[\ External $W$-emission]
		{
			\includegraphics[width=0.145\textwidth]{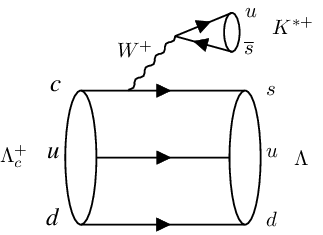}
		}
		\subfigure[\ Internal $W$-emission]
		{
			\includegraphics[width=0.145\textwidth]{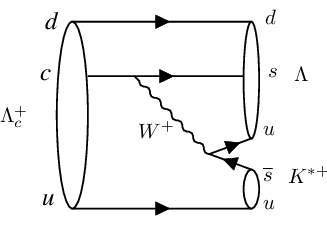}
		}
		\subfigure[\ $W$-exchange]
		{
			\includegraphics[width=0.145\textwidth]{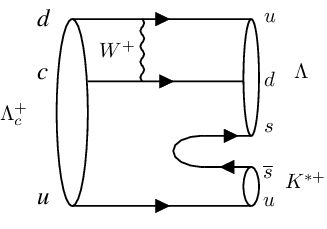}
		}
		\caption{Topology diagrams for $\Lambda_{c}^{+}\to \Lambda K^{*+}$.}
		\label{fig:feyaman_lmdkst}
	\end{figure}

	The topology diagrams of $\lmdksk$, $\lmdkspi$ and $\lmdkst$ are shown in Figs.~\ref{fig:feyaman_lmdksk} to \ref{fig:feyaman_lmdkst}. Theoretical predictions for the BFs of $\lmdksk$ and $\lmdkspi$ have been made based on $SU(3)$ flavor symmetry, with results shown in Table~\ref{tab:theoretical}.
	
	\begin{table}[!hpbt]
		\begin{center}
			\caption{Theoretical predictions for the BFs of $\Lambda_{c}^{+}\rightarrow\Lambda K_{S}^{0}K^{+}$, $\Lambda_{c}^{+}\rightarrow\Lambda K_{S}^{0}\pi^{+}$ and $\Lambda_{c}^{+}\rightarrow\Lambda K^{*+}$.}
			\begin{tabular}{ccc}
				\hline\hline
				Decay mode& C. Q. Geng~\cite{Geng:2018upx}  & Z. X. Zhao~\cite{Zhao:2018zcb}  \\ \hline
				$\Lambda_{c}^{+}\rightarrow\Lambda K_{S}^{0}K^{+}$&$(2.8\pm0.6)\times10^{-3}$&- \\
				$\Lambda_{c}^{+}\rightarrow\Lambda K_{S}^{0}\pi^{+}$&  $(4.4\pm0.7)\times10^{-3}$  &- \\
				$\Lambda_{c}^{+}\rightarrow\Lambda K^{*+}$&-&$1.97\times10^{-3}$  \\
				\hline\hline
			\end{tabular}
			\label{tab:theoretical}
		\end{center}
	\end{table}
	
	
	
	In this paper we report an improved measurement of the BF of the Cabibbo-favored decay $\lmdksk$, and the first search for the singly Cabibbo-suppressed decays $\lmdkspi$ and $\lmdkst$.
	Charge-conjugate modes are always implied throughout this paper. This analysis is performed based on electron-positron annihilation data collected by the BESIII detector at seven center-of-mass~(CM) energies ranging from $4599.53\mev$ to $4698.82\mev$, which corresponds to an integrated luminosity of 4.5 fb$^{-1}$ \cite{BESIII:energy1,BESIII:2022ulv,Ke:2023qzc}, as listed in Table~\ref{tab:luminosity}.
	
	\section{BESIII Detector and Monte Carlo Simulation}
	The BESIII detector~\cite{Ablikim:2009aa} records symmetric $e^+e^-$ collisions provided by the BEPCII storage ring~\cite{Yu:IPAC2016-TUYA01} in the CM energy range from 1.84 to 4.95~GeV, with a peak luminosity of $1 \times 10^{33}\;\text{cm}^{-2}\text{s}^{-1}$ achieved at $E_{\rm cm} = 3.78\;\text{GeV}$. BESIII has collected large data samples in this energy region~\cite{Ablikim:2019hff}. The cylindrical core of the BESIII detector covers 93\% of the full solid angle and consists of a helium-based multilayer drift chamber~(MDC), a plastic scintillator time-of-flight system~(TOF), and a CsI(Tl) electromagnetic calorimeter~(EMC), which are all enclosed in a superconducting solenoidal magnet providing a 1.0~T magnetic field. The solenoid is supported by an octagonal flux-return yoke with resistive plate counter muon identification modules interleaved with steel. The charged-particle momentum resolution at $1~{\rm GeV}/c$ is $0.5\%$, and the ${\rm d}E/{\rm d}x$ resolution is $6\%$ for electrons from Bhabha scattering. The EMC measures photon energies with a resolution of $2.5\%$ ($5\%$) at $1$~GeV in the barrel (end-cap) region. The time resolution in the TOF barrel region is 68~ps, while that in the end-cap region was 110~ps. The end-cap TOF system was upgraded in 2015 using multigap resistive plate chamber technology, providing a time resolution of 60~ps,
	which benefits 87\% of the data used in this analysis~\cite{etof}.
	
	\begin{table}[!hpbt]
		\begin{center}
			\caption{The CM energies and corresponding integrated luminosities of the analyzed data samples.}
			\begin{tabular}{cr@{.}l}
				\hline\hline
				$E_{\rm cm}$ (MeV)	&  \multicolumn{2}{c}{${\mathcal L}$ (pb$^{-1}$)} \\ \hline
				$4599.53\pm0.07\pm0.74$   &   $586.9\pm0$&$1\pm3.9$  \\
				$4611.86\pm0.12\pm0.32$   &   $103.8\pm0$&$1\pm0.6$  \\
				$4628.00\pm0.06\pm0.32$   &   $521.5\pm0$&$1\pm2.8$  \\
				$4640.91\pm0.06\pm0.38$   &   $552.4\pm0$&$1\pm2.9$  \\
				$4661.24\pm0.06\pm0.29$   &   $529.6\pm0$&$1\pm2.8$  \\
				$4681.92\pm0.08\pm0.29$   &   $1669.3\pm0$&$2\pm8.9$ \\
				$4698.82\pm0.10\pm0.39$   &   $536.4\pm0$&$1\pm2.8$  \\
				\hline \hline
			\end{tabular}
			\label{tab:luminosity}
		\end{center}
	\end{table}
	
	
	Large Monte Carlo (MC) samples are produced to simulate the annihilation of $\ee$, the initial-state radiation (ISR) effect, and the beam-energy spread using the \textsc{kkmc} generator~\cite{Jadach:2000ir}. The geometry of the BESIII detector and the interactions of charged particles and photons are simulated by a \textsc{geant4}-based detector simulation package~\cite{geant4}.
	The MC samples consist of pair production of $\lcp\lcm$ , open-charm mesons, ISR processes to lower-mass $\psi$ states, and the continuum processes $e^{+}e^{-}\rightarrow q\bar{q}$ ($q=u,d,s$). The known decay modes of charmed hadrons and charmonium states are modeled using \textsc{evtgen}~\cite{Lange:2001uf, Ping:2008zz} with BFs taken from the Particle Data Group (PDG)~\cite{pdg2022}. The remaining unknown decays are modeled with \textsc{lundcharm}~\cite{Chen:2000tv}.
	Additionally, exclusive signal PHSP MC samples are generated to describe the decays of $\lmdksk$, $\lmdkspi$, and $\lmdkst$, to determine the detection efficiencies.

	
	\section{EVENT SELECTION}
	Each of the three signal modes contains five charged particles in the final states, which must be reconstructed as tracks  in the MDC.  All tracks except for those from $\Ks$ and $\Lambda$ decays are required to have a closest approach of less than 1~cm in the transverse plane with respect to the interaction point (IP) and less than 10~cm along the positron beam direction. The polar angle $\theta$ with respect to the symmetry axis of the MDC is required to satisfy $|\!\cos\theta|<0.93$.
	The likelihoods $\mathcal{L}$ under $\pi, K$ and $p$ hypotheses are assigned by combining the information from the TOF and the specific ionization energy loss (d$E/$d$x$) in the MDC. A charged track is identified as a $\pi$ or $K$ if $\mathcal{L}(\pi)>\mathcal{L}(K)$ and $\mathcal{L}(K)>\mathcal{L}(\pi)$, respectively.
	
	Candidates for $K_{S}^{0}$ and $\Lambda$ hadrons are formed by combining two oppositely charged tracks into the final states $\pi^{+}\pi^{-}$ and $p\pi^{-}$.
	For these two tracks, the distances of closest approaches to the IP must be within $\pm$20 cm along the beam direction, while there is no requirement for the constraint perpendicular to the beam direction. The charged daughter pion is not subjected to the particle identification (PID) requirements described above, while the PID for proton candidate from the $\Lambda$ decay is required to satisfy $\mathcal{L}(p)>\mathcal{L}(K)$ and $\mathcal{L}(p)>\mathcal{L}(\pi)$ to improve the signal significance.
	The two daughter tracks are constrained to originate from a common decay vertex by requiring the $\chi^2$ of the vertex fit to be less than 100.
	Furthermore, the decay vertex is required to be separated from the IP by a distance of at least twice the fitted vertex resolution.
	The fitted momenta of the $\pi^{+}\pi^{-}$ and $p\pi^{-}$ pairs are used in the subsequent analysis.
	The $p\pim$ combination with invariant mass lying within $[1090,1140]\,\mevcc$ and the $\pip\pim$ combination with invariant mass lying within $[450,540]\,\mevcc$ are selected as $\Lambda$ and $\Ks$ candidates, respectively.
	
	The $\lcp$ candidates are formed by combining all the $\Lambda$, $\Ks$ and $\kp(\pip)$ candidates in an event. Two kinematic variables, the energy difference $\Delta{}E \equiv E - E_{\rm beam}$ and the beam-constrained mass $M_{\rm BC}\equiv \sqrt{E_{\rm beam}^2/c^4-|\vec{p}|^2/c^2}$, are used to isolate the $\lcp$ candidates in the subsequent analysis, where $E_{\rm beam}$ is the average value of the $e^+$ and $e^-$ beam energies and $\vec{p}$ is the measured momentum of $\lcp$ in the laboratory system. All the candidates are required to be within $-0.02\gev<\Delta{}E<0.02\gev$. If more than one candidate in an event satisfies all the above requirements, the one with the lowest $|\Delta{}E|$ is selected.
	
	\section{ANALYSIS}
	The BF of each signal decay is calculated by
	\begin{eqnarray}
	\begin{aligned}
	\mathcal{B}_{\rm sig} = \frac{N}{2\cdot\mathcal{B}_{\rm int}\cdot\sum_{i}\left(  N_{\lambdacp\lambdacm}^{i}\cdot\varepsilon_{i}\right) },
	\end{aligned}
	\label{eq:bf}
	\end{eqnarray}
	where $N$ is the signal yield obtained from data combined from all energy points, $N_{\lcp\lcm}^{i}$ is the total number of $\lcp\lcm$ pairs produced in data \cite{BESIII:2022ulv,BESIII:2023rwv}, $\varepsilon_{i}$ is the detection efficiency, and $i$ denotes the $i$-th energy point. $\mathcal{B}_{\rm int}$ is the product BF of the intermediate states $\Lambda$, $\Ks$ (and $\kst$ for $\lmdkst$).
	
\begin{figure}[!hpbt]
		\centering
		{
			\includegraphics[width=0.48\textwidth]{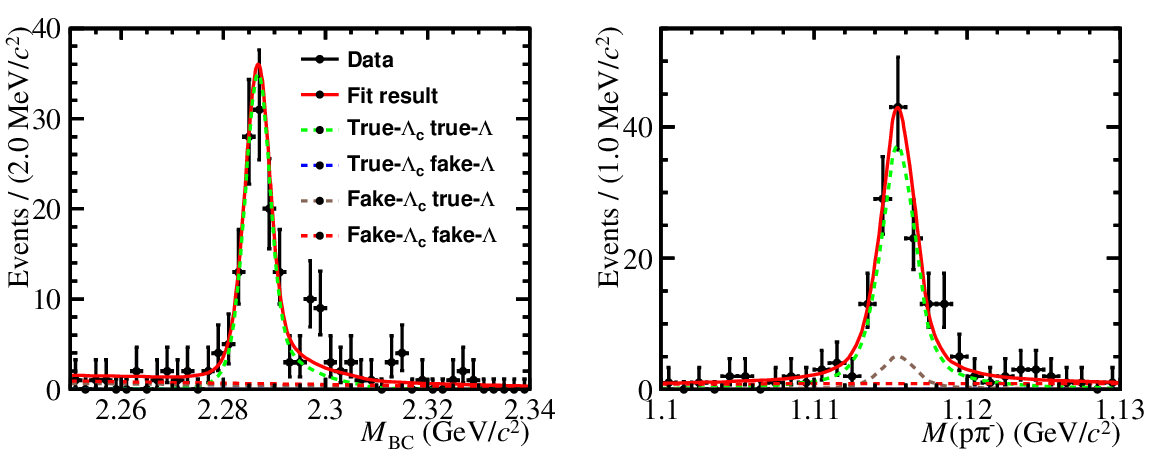}

			\includegraphics[width=0.48\textwidth]{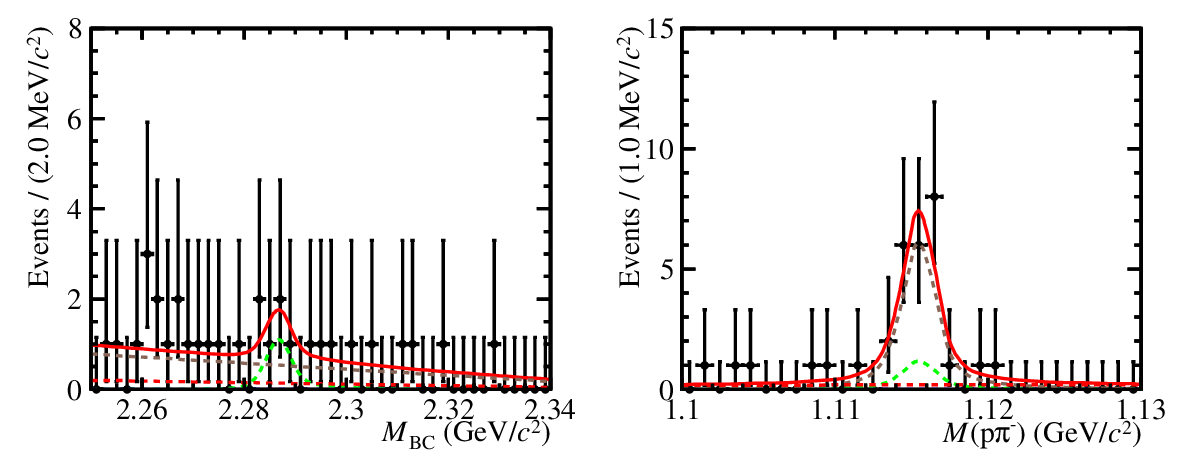}
		}
	\vspace{-2em}
		\caption{The 2-D simultaneous fit result projection on the $M_{\rm BC}$ and $\mppi$ invariant-mass distributions of the $\lmdksk$ candidates in the $\Ks$ signal (top row) and sideband (bottom row) regions. 
}
		\label{fig:fitlmdksk_sigsb}
	\end{figure}

	For $\lmdksk$, the signal yield is obtained through a two-dimensional~(2-D) extended unbinned maximum likelihood fit on the $M_{\rm BC}$ and $\mppi$ invariant-mass distributions, as shown in Fig.~\ref{fig:fitlmdksk_sigsb}. To estimate the background from $\Ks$ candidates from incorrect pion combinations, the fit is performed simultaneously for the samples in the $\Ks$ signal and sideband regions, which are defined as $[0.487,0.511]\gevcc$ and $[0.450,0.470]\cup[0.520,0.540]\gevcc$, respectively. The signals are described by MC simulated shapes convolved with Gaussian functions, while the backgrounds are modeled by  linear functions. The shapes are shared in the fits for the $\Ks$ signal and sideband regions.
	
\begin{figure}[!hpbt]
		\centering
		{
			\includegraphics[width=0.48\textwidth]{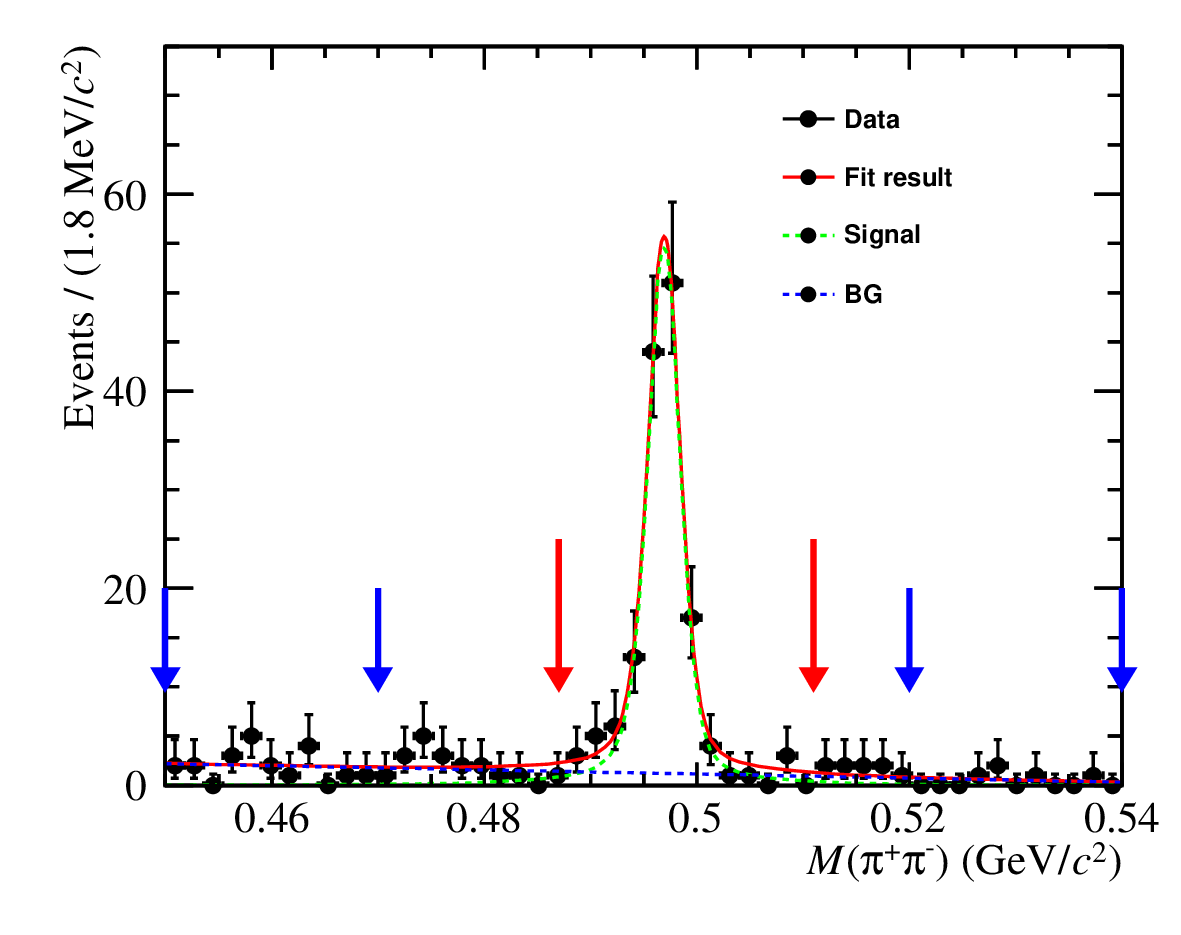}
		}
		\vspace{-2em}
		\caption{The one-dimensional fit to the $M(\pi^{+}\pi^{-})$ distribution for the $\Lambda_{c}^{+}\rightarrow \Lambda \Ks K^{+}$ candidates. 
The red arrows indicate the signal region and the blue arrows indicate the sideband regions.
}
		\label{fig:fitlmdksk_ks}
	\end{figure}

	The ratio of fake-$\Ks$ background $f_{\Ks}$ between the $\Ks$ signal and sideband regions is determined to be $0.56 \pm 0.01$ from a one-dimensional fit to the $\mpipi$ distribution, as shown in Fig.~\ref{fig:fitlmdksk_ks}.

	The signal yield $N^{\Lambda\Ks\kp}$, after subtracting the combinatorial background as estimated from the $\Ks$ sideband region, is calculated to be $128.9\pm12.7$ by
	\begin{eqnarray}
	\begin{aligned}
	N^{\Lambda\Ks K^+}=N_{\rm sig}^{\Lambda\Ks K^+}-f_{\Ks}\cdot N_{\rm sb}^{\Lambda\Ks K^+}.
	\end{aligned}
	\label{eq:nsig1}
	\end{eqnarray}
\noindent where the `sig' and `sb' subscripts refer to the measurements in the signal and sideband regions, respectively.
	
\begin{figure}[!hpbt]
		\centering
		{
			\includegraphics[width=0.48\textwidth]{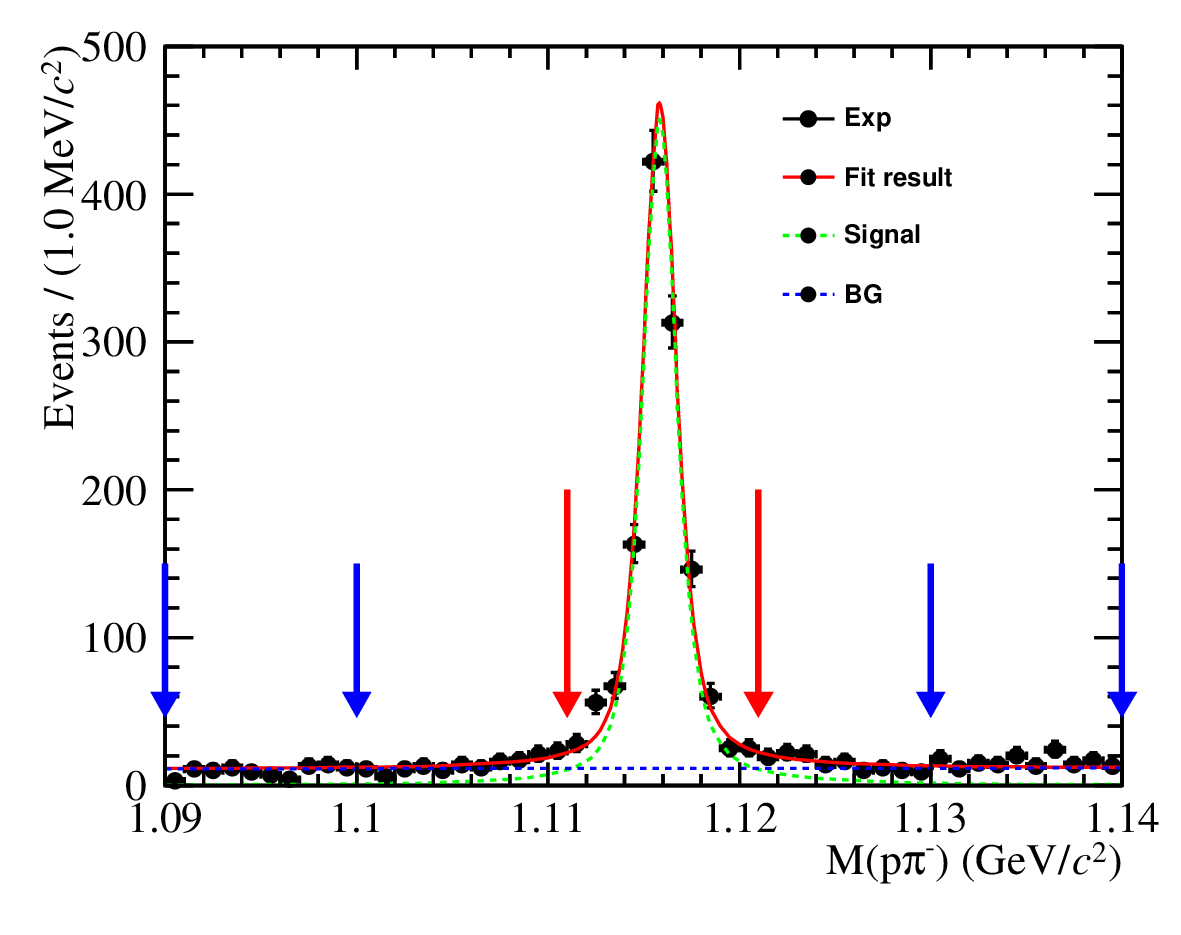}
		}
		\vspace{-2em}
		\caption{The one-dimensional fit to the $M(p\pi^{-})$ distribution for $\Lambda_{c}^{+}\rightarrow \Lambda K^{*+}$ candidates. 
The red arrows indicate the signal region and the blue arrows indicate the sideband regions.
}
		\label{fig:fitlmdkst_lmd}
	\end{figure}

	For $\lmdkspi$($\lmdkst$), a clear peak is found around the known $\kst$ mass in the distribution of $\mkspi$. However, due to the limited sample size, a partial-wave analysis is not feasible. To obtain the signal yield, a 3-D extended unbinned maximum likelihood fit on the distributions of $M_{\rm BC}$, $\mpipi$ and $\mkspi$ is performed simultaneously in the $\Lambda$ signal and sideband regions, which are defined as $[1.111,1.121]\gevcc$ and $[1.090,1.100]\cup[1.130,1.140]\gevcc$, respectively. The signal yields of the non-resonant~(NR) $\lmdkspi$, $\lmdkst$ and total $\lmdkspi$ are determined via
	
	\begin{eqnarray}
	\begin{aligned}
	N^{NR}=&N_{\rm sig}^{NR}-f_{\Lambda}\cdot N_{\rm sb}^{NR},\\
	N^{\Lambda\kst}=&N_{\rm sig}^{\Lambda\kst}-f_{\Lambda}\cdot N_{\rm sb}^{\Lambda\kst},\\
	N^{\Lambda\Ks\pip}=&N^{NR}+N^{\Lambda\kst}+N^{\rm int}, \\
	\label{eq:nsig2_inter}
	\end{aligned}
	\end{eqnarray}
	where $N_{\rm sig(sb)}^{NR/\Lambda\kst}$ is the yield in the $\Lambda$ signal (sideband) region for the NR or $\Lambda\kst$ component. The $f_{\Lambda}$ is the ratio of backgrounds in the $\Lambda$ signal and sideband regions, and is estimated to be $0.50\pm0.01$ from a one-dimensional fit on the $\mppi$ distribution, as shown in Fig.~\ref{fig:fitlmdkst_lmd}. The $N^{\rm int}$ is the signal yield of the interference term between the NR and $\Lambda\kst$ components. In the fit, $N^{\Lambda\kst}$, $N^{\Lambda\Ks\pip}$, $N_{\rm sb}^{NR}$, $N_{\rm sb}^{\Lambda\kst}$ are free parameters.
	
\begin{figure*}[!hpbt]
		\centering
		{
			\includegraphics[width=1\textwidth]{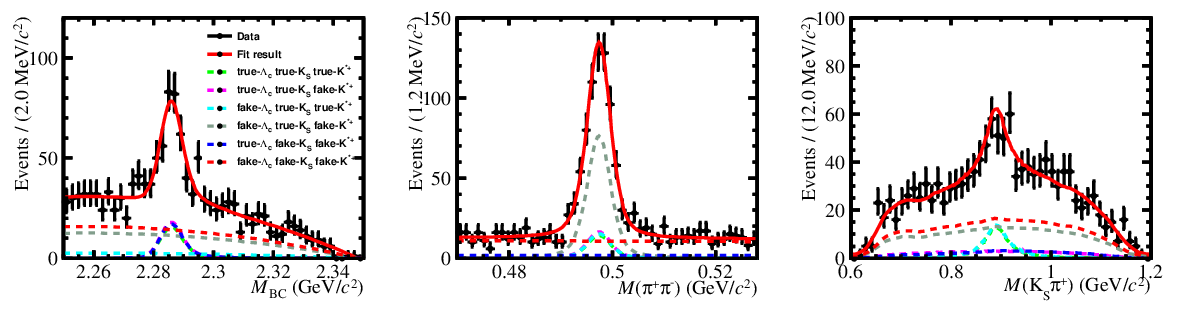}
			\includegraphics[width=1\textwidth]{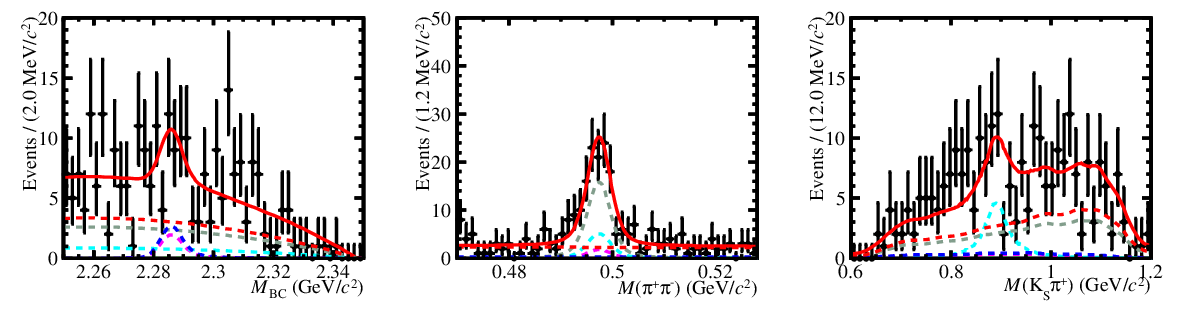}
		}
		\vspace{-2em}
		\caption{The 3-D simultaneous fit result projection on the distributions of $M_{\rm BC}$, $\mpipi$ and $\mkspi$ for $\lmdkspi$ and $\lmdkst$ candidates in the $\Lambda$ signal (top row) and sideband (bottom row) regions. 
}
		\label{fig:fitlmdkst_sigsb}
	\end{figure*}

	Initially, the fit is performed assuming no interference between the NR and $\Lambda\kst$($N^{\rm int}=0$) components, as shown in Fig.~\ref{fig:fitlmdkst_sigsb}. The signals are described by the MC simulated shapes convolved with Gaussian functions that account for differences in resolution between the MC simulation and data. The backgrounds are modeled by 2nd-order Chebyshev polynomial functions in the $M_{\rm BC}$ distribution, linear functions in the $\mpipi$ distribution, and MC-simulated shapes in the $\mkspi$ distribution. The measured signal yields are $N^{\Lambda\Ks\pip}=167\pm25$ and $N^{\Lambda\kst}=80\pm19$.
	
\begin{figure}[!hpbt]
		\centering
		{
			\includegraphics[width=0.48\textwidth]{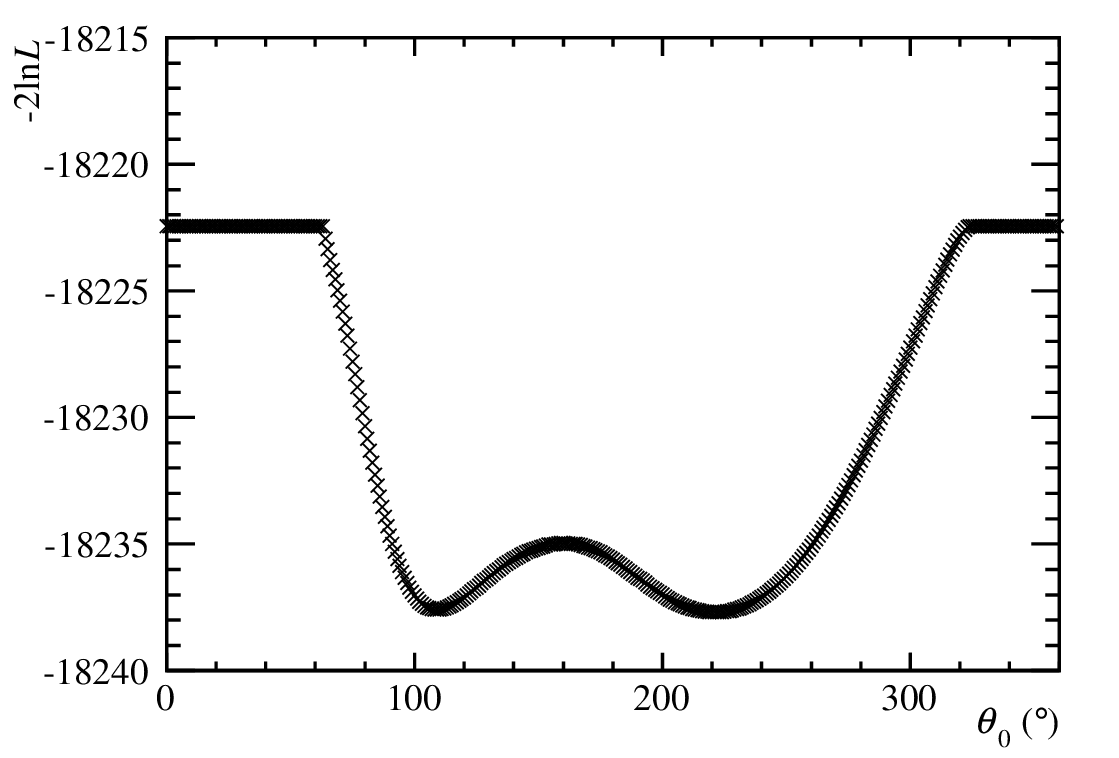}
		}
		\vspace{-2em}
		\caption{The distribution of $-2{\rm ln}L$ of the fit results in the range  $0^{\circ}\leq\theta_{0}<360^{\circ}$.}
		\label{fig:fcn}
	\end{figure}

\begin{figure*}[!hpbt]
\centering
{
	\includegraphics[width=1.0\textwidth]{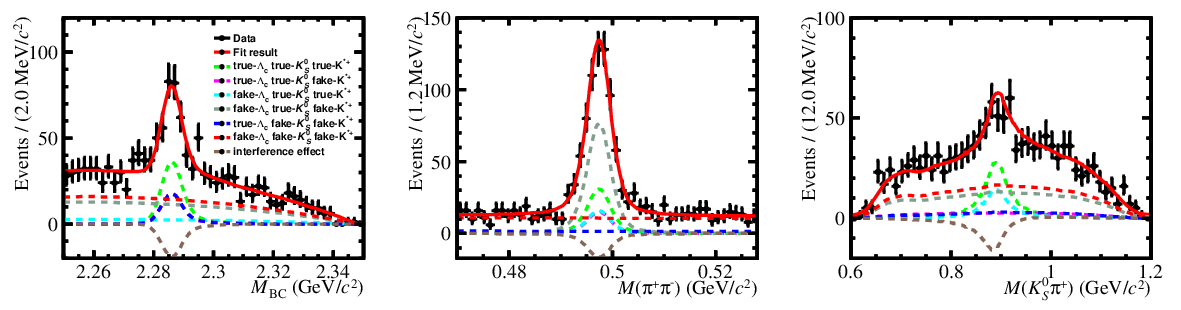}
	\includegraphics[width=1.0\textwidth]{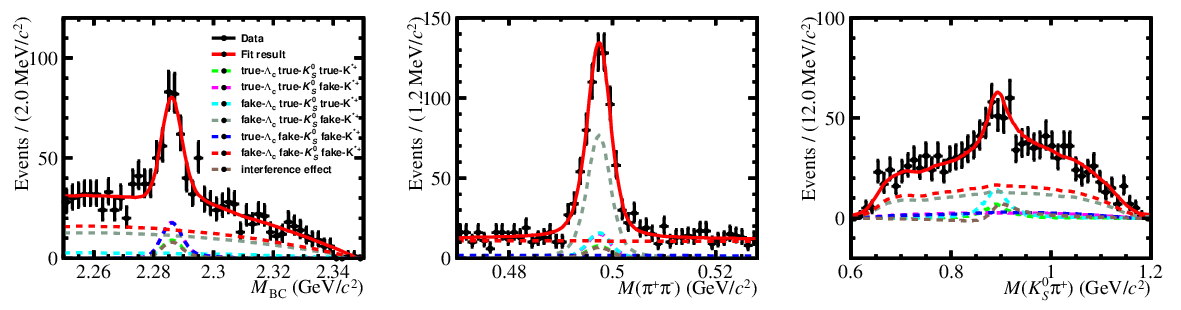}
}
	\vspace{-2em}
\caption{The 3-D simultaneous fit result projection on the distributions of $M_{\rm BC}$, $M(\pip\pim)$ and $M(\Ks\pi^{+})$ for $\lambdacp\to \Lambda \Ks \pip$ and $\lambdacp\to \Lambda \kst$ in the $\Lambda$ signal region under the assumptions of (top row) $\theta=109^\circ$ and (bottom row) $\theta=221^\circ$. 
}
\label{fig:fitlmdkst_inter_sigsb}
\end{figure*}

	However, since the width of $\kst$ is relatively broad, interference effects cannot be neglected. Therefore, 3-D fits including $\mkspi$ are performed under different interference assumptions for the $\lmdkspi$ decay. These assumptions are described by the relative phase angle $\theta_{0}$ between the $\lmdkst$ and NR processes. The one-dimensional probability density functions (PDFs) of the $\lmdkst$ and NR components are denoted as $f^{\Lambda\kst}$ and $f^{NR}$, respectively, and are also constructed with no interference. The interference term in the PDF, $f^{\rm int}$, and yield $N^{\rm int}$ are expressed as a function of $f^{\Lambda\kst}$ and $f^{NR}$ and a function of $N^{NR}$ and $N^{\Lambda\kst}$ via
	
	\begin{eqnarray}
		\begin{aligned}
			f^{\rm int}(M)=&2a\cos(\theta(M)+\theta_{0})\\
			&\cdot\sqrt{f^{\Lambda\kst}(M)\cdot f^{NR}(M)},
			\label{eq:pdf_inter}
		\end{aligned}
	\end{eqnarray}
	\begin{eqnarray}
	\begin{aligned}
		N^{\rm int}=\frac{1}{a}\sqrt{N^{NR}\cdot N^{\Lambda\kst}},\\
		\end{aligned}
		\label{eq:nsig_inter}
	\end{eqnarray}
	where $a$ is the normalization factor, and $\theta(M)$ is the phase angle of $\lmdkst$, calculated from the Breit-Wigner function:
	\begin{eqnarray}
		|\sqrt{f^{\Lambda\kst}(M)}|e^{i\theta(M)}=&BW(M)\\
		=&\frac{1}{m_{0}^{2}-M^{2}-im_{0}\Gamma_{0}},
	\end{eqnarray}
	\begin{eqnarray}
		\theta(M)=\arccos\frac{m_0^2-M^2}{\sqrt{(m_0^2-M^2)^2+m_0^2\cdot\Gamma_0^2}}.
	\end{eqnarray}
Here, $m_0$ and $\Gamma_0$ are the known mass and decay width of the $K^{*+}$, respectively, taken from the PDG values~\cite{pdg2022}. The value of $\theta_0$ is unknown, and thus a series of 3-D simultaneous fits are performed to determine the BFs with different $\theta_0$ in the range of $0^{\circ}\leq\theta_0<360^{\circ}$ with a step of $1^{\circ}$. The distribution of $-2{\rm ln}\mathcal{L}$ for the fits is shown in Fig.~\ref{fig:fcn}.
	It reaches a minimal when $\theta_0$ takes a value $221^{\circ}$ or $109^{\circ}$, with the corresponding fit results shown in Fig.~\ref{fig:fitlmdkst_inter_sigsb}.
	From these fits, the signal yields of $\lmdkspi$ are $N^{\Lambda\Ks\pip}(\theta_0=109^{\circ})=161\pm22$ and $N^{\Lambda\Ks\pip}(\theta_0=221^{\circ})=162\pm24$, and the signal yields of $\lmdkst$ are $N^{\Lambda\kst}(\theta_0=109^{\circ})=173\pm34$ and $N^{\Lambda\kst}(\theta_0=221^{\circ})=43\pm15$.
	
	The statistical significance, shown in Table~\ref{tab:bf}, is calculated with $\sqrt{-2\rm ln(\mathcal{L}_{0}^{\rm stat}/\mathcal{L}_{\rm max}^{\rm stat})}$, where $\mathcal{L}_{0}^{\rm stat}$ and $\mathcal{L}_{\rm max}^{\rm stat}$ are the maximum likelihood with and without signal. We observe the $\lmdkspi$ decay for the first time with statistical significance of $8.9\sigma$, and we find  evidence for $\lmdkst$ with a statistical significance of $4.7\sigma$.
	
	The signal MC samples are used to obtain the detection efficiency. The efficiencies of $\lmdksk$ and $\lmdkst$ are determined with the phase space~(PHSP) MC samples directly. The efficiency of $\lmdkspi$ is obtained by
	\begin{eqnarray}
		\varepsilon_{i}=\frac{\varepsilon^{\alpha}_{i}\varepsilon^{\beta}_{i}(N^{NR}+N^{\Lambda \kst})}{\varepsilon^{\beta}_{i}N^{NR}+\varepsilon^{\alpha}_{i}N^{\Lambda \kst}},
	\end{eqnarray}
	where $N^{NR(\Lambda\kst)}$ is the signal yield when ignoring interference, $\varepsilon_{i}^{\alpha}$ and $\varepsilon_{i}^{\beta}$ are the detection efficiencies for the non-resonant $\lmdkspi$ and $\lmdkst$, respectively. The detection efficiencies for each decay mode at different energy points are shown in Table~\ref{tab:efficiency}.

	\begin{table}[!hpbt]
		\begin{center}
			\caption{The signal yields, BFs and significance for each decay mode.}
			\begin{tabular}{cccc}
				\hline \hline
				Decay mode  &$N$&  $\mathcal{B}$~($\times10^{-3}$) & Significance\\ \hline
				$\Lambda \Ks K^{+}$ &$128.9\pm12.7$& $3.04\pm0.30$  & $10.6\sigma$\\ \hline
				$\Lambda \Ks \pi^{+}$ &$166.5\pm25.3$& $1.73\pm0.27$  & \multirow{3}{*}{$8.9\sigma$}\\
				$\Lambda \Ks \pi^{+}(\theta_{0}=109^{\circ})$ &$161.0\pm21.9$& $1.73\pm0.23$  & \\
				$\Lambda \Ks \pi^{+}(\theta_{0}=221^{\circ})$ &$161.5\pm23.7$& $1.73\pm0.25$  & \\ \hline
				$\Lambda K^{*+}$ &$79.7\pm19.2$& $2.40\pm0.58$ & \multirow{3}{*}{$4.7\sigma$}\\
				$\Lambda K^{*+}(\theta_{0}=109^{\circ})$ &$172.9\pm23.6$& $5.21\pm0.71$ & \\
				$\Lambda K^{*+}(\theta_{0}=221^{\circ})$ &$42.9\pm14.7$& $1.29\pm0.44$ & \\\hline\hline
			\end{tabular}
			\label{tab:bf}
		\end{center}
	\end{table}

	\begin{table}[!hpbt]
		\begin{center}
			\caption{Detection efficiencies  (in \%) for each decay mode at different CM energy points.}
			\begin{tabular}{cccc}
				\hline\hline
				$E_{\rm cm}~(\rm MeV)$&$\lmdksk$&$\lmdkspi$&$\lmdkst$\\
				\hline
				$4599.53$&$6.56\pm0.02$&$16.53\pm0.04$&$17.24\pm0.04$\\
				$4611.86$&$5.92\pm0.02$&$14.80\pm0.04$&$15.46\pm0.04$\\
				$4628.00$&$5.91\pm0.02$&$14.39\pm0.04$&$15.01\pm0.04$\\
				$4640.91$&$6.09\pm0.02$&$14.28\pm0.03$&$14.78\pm0.04$\\
				$4661.24$&$6.23\pm0.02$&$13.94\pm0.03$&$14.35\pm0.04$\\
				$4681.92$&$6.34\pm0.02$&$13.61\pm0.03$&$14.03\pm0.03$\\
				$4698.82$&$6.38\pm0.02$&$13.35\pm0.03$&$13.74\pm0.03$\\
				\hline\hline
			\end{tabular}
			\label{tab:efficiency}
		\end{center}
	\end{table}

	
	\section{Systematic  Uncertainties}
	
	The uncertainties related to the efficiencies of both the PID and tracking of the charged tracks are assigned as 1\% per track, respectively, based on a study of a control sample of $\ee\to\kp\km\pip\pim$ events~\cite{BESIII:2019kfh}.
	The uncertainties associated with the reconstruction of $\Lambda$ and $\Ks$ decays have been studied in Ref.~\cite{BESIII:2015bjk} and Ref.~\cite{BESIII:2015jmz}, respectively, and are assigned as $2.5\%$ and $1.5\%$ in this analysis.
	The uncertainties in the BFs of the intermediate states of $\Lambda \to p\pi^{-}$ and $\Ks \to \pi^{+}\pi^{-}$ are taken from the PDG~\cite{pdg2022}, and are $0.8\%$ for $\Lambda \to p\pi^{-}$ and $0.1\%$ for $\Ks \to \pi^{+}\pi^{-}$.
	The values for $N_{\lambdacp\lambdacm}$ at each energy point are taken from Ref.~\cite{BESIII:2022ulv,BESIII:2023rwv}. The associated uncertainties are  $3.1\%$ for $\lambdacp\to \Lambda \Ks K^{+}$ and $2.8\%$ for $\lambdacp\to \Lambda \Ks \pi^{+}$($\lambdacp\to \Lambda K^{*+}$).
	The impact of the uncertainties $f_{\Ks/\Lambda}$ on the measured BFs is negligible. 
	
	The uncertainty associated with the efficiency of the 2-D or 3-D fit is estimated by varying the signal  and background shapes. The uncertainty due to signal shape is assessed by replacing the smeared-Gaussian resolution function with a double-Gaussian function. The difference in the derived BF from the two approaches is taken as the systematic uncertainty. This is $2.6\times10^{-5}$ for $\lambdacp\to \Lambda \Ks K^{+}$, $2.4\%$ for $\lambdacp\to \Lambda \Ks \pi^{+}$, and $0.7\%$ for $\lambdacp\to \Lambda K^{*+}$. To estimate the uncertainty arising from the choice of background parametrization, we change the background shape to a polynomial function with fixed parameters obtained from the fit to the background MC samples. The difference in the BF is taken as the uncertainty. This is $0.1\%$ for $\lambdacp\to \Lambda \Ks K^{+}$, $2.7\%$ for $\lambdacp\to \Lambda \Ks \pi^{+}$, and $0.3\%$ for $\lambdacp\to \Lambda K^{*+}$.   The overall systematic uncertainty from the 2-D (3-D) fit is taken to be the sum in quadrature of these two contributions, which is $0.1\%$ for $\lambdacp\to \Lambda \Ks K^{+}$, and $3.6\%$ for $\lambdacp\to \Lambda \Ks \pi^{+}$, $0.8\%$ for $\lambdacp\to \Lambda K^{*+}$.
	
	To estimate the uncertainty due to the $\Delta E$ requirement,  we convolve a Gaussian function with the shape found in MC, the parameters of which we fit on data.  This function accounts for differences in resolution between data and MC.  We then remeasure the efficiency on MC with this modified resolution, and take the  observed changes in the BFs  as the uncertainties, which are $0.0015\%$ for $\lambdacp\to \Lambda \Ks K^{+}$, $0.5\%$ for $\lambdacp\to \Lambda \Ks \pi^{+}$, and $0.4\%$ for $\lambdacp\to \Lambda K^{*+}$.
	
	The uncertainty due to the MC sample size is calculated by
	\begin{eqnarray}
		\begin{aligned}
			\frac{\Delta\varepsilon}{\varepsilon}=\frac{\sqrt{\sum_{i}\left[ N_{(\lambdacp\lambdacm)_{i}}\cdot \Delta\varepsilon_{i}\right]^{2} }}{\sum_{i}\left[ N_{(\lambdacp\lambdacm)_{i}}\cdot \varepsilon_{i}\right]},
		\end{aligned}
	\end{eqnarray}
	where $\varepsilon_{i}$ and $N_{(\lambdacp\lambdacm)_{i}}$ is the efficiency and the number of $\lambdacp$ pairs at the $i$-th energy point. These uncertainties are $0.2\%$ for $\lambdacp\to \Lambda \Ks K^{+}$, $0.1\%$ for $\lambdacp\to \Lambda K^{*+}$, and $0.1\%$ for $\lambdacp\to \Lambda \Ks \pi^{+}$.
	
	The PHSP MC model is used as the baseline model in the measurement. An alternative choice is to reweight the PHSP MC based on the background-subtracted data. The difference in efficiency between these two models is then assigned as  the associated uncertainty. This is $2.6\%$ for $\lambdacp\to \Lambda \Ks K^{+}$, $1.6\%$ for $\lambdacp\to \Lambda K^{*+}$, and $0.8\%$ for $\lambdacp\to \Lambda \Ks \pip$.
	
	The total systematic uncertainty is taken to be the sum in quadrature of the above contributions, which are assumed to be uncorrelated, and is shown in Table~\ref{tab:sys_err} for each decay mode.
	The overall significance of the $\lmdkst$ signal, after smearing the likelihood curve with the systematic uncertainty, is $4.66\sigma$.  
	
	\begin{table}[!hpbt]
		\begin{center}
			\caption{Relative systematic uncertainties (in \%) in the BF measurements, where `-' indicates the uncertainty is negligible.}
			\begin{tabular}{lccc}
				\hline \hline
				Source            &  \makecell{$\Lambda \Ks K^{+}$} & \makecell{$\Lambda K^{*+}$}&  \makecell{$\Lambda \Ks \pi^{+}$} \\
				\hline
				PID 								&1.0	&1.0	&1.0	\\
				Tracking         					&1.0	&1.0	&1.0	\\
				$\Lambda$ reconstruction			&2.5	&2.5	&2.5	\\
				$\Ks$ reconstruction				&1.5	&1.5	&1.5 	\\
				$\mathcal{B}_{\rm int}$					&0.8	&0.8	&0.8	\\
				$N_{\lambdacp\lambdacm}$			&3.1	&2.8	&2.8	\\
				$f_{\Ks/\Lambda}$					&-	 	&-		&-		\\
				2-D/3-D fit						&-		&0.8	&3.6	\\
				$\Delta E$							&-		&0.4	&0.5	\\
				MC sample size						&0.2	&0.1	&0.1	\\
				MC model							&2.6	&1.6	&0.8	\\
				\hline
				Total             					&5.3	&4.7	&5.7	\\
				\hline \hline
			\end{tabular}
			\label{tab:sys_err}
		\end{center}
	\end{table}

	
	\section{Summary}
	By analyzing $e^+e^-$ collision data corresponding to an integrated luminosity of $4.5\,\mathrm{fb}^{-1}$ taken in the CM energy range from $4599.53\mev$ to $4698.82\mev$ with the BESIII detector, we measure the BFs of $\Lambda_{c}^{+}\rightarrow\Lambda K_{S}^{0}K^{+}$, $\Lambda_{c}^{+}\rightarrow\Lambda K_{S}^{0}\pi^{+}$ and $\Lambda_{c}^{+}\rightarrow\Lambda K^{*+}$. The obtained results are shown in Table~\ref{tab:comparelmdkspilmdksts0kspi}. The BF of $\lmdksk$ is measured to be $(3.04\pm0.30\pm0.16)\times 10^{-3}$, which is consistent with the PDG value but with improved precision~\cite{pdg2022}. The singly Cabibbo-suppressed decay $\lmdkspi$ is observed for the first time and its decay BF is measured to be $(1.73\pm0.26\pm0.10)\times 10^{-3}$, which is about $4\sigma$ lower than the predictions based on SU(3) flavor symmetry~\cite{Geng:2018upx}. A similar discrepancy is  observed in the $\lcp\to\Lambda K^{+}$ decay~\cite{BESIII:lc2lmdk}. These discrepancies indicate that more intensive investigations are needed to better understand $\Lambda_c^+$ decays involving a $\Lambda$ with one strange hadron. The intermediate decay $\lmdkst$ is studied for the first time and considered under different interference assumptions. Its decay BF is determined to be $(2.40\pm0.58\pm0.11)\times 10^{-3}$ ignoring interference effect, $(5.21\pm0.71\pm0.25)\times 10^{-3}$ for $\theta_{0}=109^{\circ}$, and $(1.29\pm0.44\pm0.06)\times 10^{-3}$ for $\theta_{0}=221^{\circ}$. All these measurement are statistically dominated. With the larger data sets which are foreseen to be collected near the $\lcp\lcm$ threshold in the coming years~\cite{BESIII:2020nme}, it will be possible to obtain more precise results concerning the decay mechanisms of charmed baryons.

	\begin{table}[!hpbt]
		\begin{center}
			\caption{The comparison of the measured BFs (in $10^{-3}$) with the PDG average and theoretical calculations.}
			\begin{tabular}{cccc}
				\hline\hline
				Decay mode&\makecell{PDG~\cite{pdg2022}}& \makecell{Theory\\~\cite{Geng:2018upx}~\cite{Zhao:2018zcb}}  &This work  \\ \hline
				$\Lambda K_{S}^{0}K^{+}$&$2.85\pm0.55$&$2.8\pm0.6$&$3.04\pm0.30\pm0.16$ \\\hline
				$\Lambda K_{S}^{0}\pi^{+}$&-&  $4.4\pm0.7$  &$1.73\pm0.26\pm0.10$ \\\hline
				\begin{tabular}[c]{@{}c@{}}$\Lambda K^{*+}$\\ (no\ interference)\end{tabular}&\multirow{3}{*}{-}&\multirow{3}{*}{$1.97$}&$2.40\pm0.58\pm0.11$  \\
				\makecell{$\Lambda K^{*+}$ $(\theta_{0}=109^{\circ})$}& & &$5.21\pm0.71\pm0.25$ \\
				\makecell{$\Lambda K^{*+}$ $(\theta_{0}=221^{\circ})$}& & &$1.29\pm0.44\pm0.06$ \\
				\hline\hline
			\end{tabular}
			\label{tab:comparelmdkspilmdksts0kspi}
		\end{center}
	\end{table}

\acknowledgments
The BESIII Collaboration thanks the staff of BEPCII and the IHEP computing center for their strong support. This work is supported in part by National Key R\&D Program of China under Contracts Nos. 2020YFA0406400, 2020YFA0406300, 2023YFA1606000 2023YFA1609400; National Natural Science Foundation of China (NSFC) under Contracts Nos. 11635010, 12105127, 11735014, 11835012, 11935015, 11935016, 11935018, 11961141012, 12025502, 12035009, 12035013, 12061131003, 12192260, 12192261, 12192262, 12192263, 12192264, 12192265, 12221005, 12225509, 12235017; the Chinese Academy of Sciences (CAS) Large-Scale Scientific Facility Program; the CAS Center for Excellence in Particle Physics (CCEPP); Joint Large-Scale Scientific Facility Funds of the NSFC and CAS under Contract No. U1832207; CAS Key Research Program of Frontier Sciences under Contracts Nos. QYZDJ-SSW-SLH003, QYZDJ-SSW-SLH040; 100 Talents Program of CAS;Fundamental Research Funds for Central Universities, Lanzhou University, University of Chinese Academy of Sciences; The Institute of Nuclear and Particle Physics (INPAC) and Shanghai Key Laboratory for Particle Physics and Cosmology; European Union's Horizon 2020 research and innovation programme under Marie Sklodowska-Curie grant agreement under Contract No. 894790; German Research Foundation DFG under Contracts Nos. 455635585, Collaborative Research Center CRC 1044, FOR5327, GRK 2149; Istituto Nazionale di Fisica Nucleare, Italy; Ministry of Development of Turkey under Contract No. DPT2006K-120470; National Research Foundation of Korea under Contract No. NRF-2022R1A2C1092335; National Science and Technology fund of Mongolia; National Science Research and Innovation Fund (NSRF) via the Program Management Unit for Human Resources \& Institutional Development, Research and Innovation of Thailand under Contract No. B16F640076; Polish National Science Centre under Contract No. 2019/35/O/ST2/02907; The Swedish Research Council; U. S. Department of Energy under Contract No. DE-FG02-05ER41374; ANID PIA/APOYO AFB230003, Chile


\end{document}